\documentclass[twocolumn,prl,aps,superbib,tightenlines,floatfix]{revtex4}

\usepackage{amsfonts}
\usepackage{amsmath}
\usepackage{amssymb}
\usepackage{graphicx}

\begin{document}

\title{Spin filtering implemented through Rashba and weak magnetic
modulations}
\author{S. J. Gong}
\affiliation{Surface Physics Laboratory (National Key Laboratory), Fudan University,
Shanghai 200433, China}
\author{Z. Q. Yang}
\thanks{To whom correspondence should be addressed}
\email[Email address: ]{zyang@fudan.edu.cn}
\affiliation{Surface Physics Laboratory (National Key Laboratory), Fudan University,
Shanghai 200433, China}

\begin{abstract}
We present two theoretical schemes for spin filters in one-dimensional
semiconductor quantum wires with spatially modulated Rashba spin-orbit
coupling (SOC) as well as weak magnetic potential. For case I, the SOC is
periodic and the weak magnetic potential is applied uniformly along the
wire. Full spin polarizations with opposite signs are obtained within two
separated energy intervals. For case II, the weak magnetic potential is
periodic while the SOC is uniform. An ideal negative/positive switching
effect for spin polarization is realized by tuning the strength of SOC. The
roles of SOC, magnetic potential, and their coupling on the spin filtering
are analyzed.

\end{abstract}

\maketitle

\textit{\ }\textbf{Introduction:} The spin filter, which can generate
spin-polarized current out of an unpolarized source, is one of the research
focuses in the field of the semiconductor spintronics. Among the various
schemes for spin filters, magnetically modulated nanostructures \cite%
{Egue,Zhai,Lu,Wang,Guo}, such as dilute magnetic semiconductor
heterostructures \cite{Egue} and two-dimensional electron gases (2DEGs)
subject to local magnetic fields \cite{Zhai}, have been the dominant
choices. Spin filtering in such schemes, however, usually requires strong
magnetic fields, which still remain a challenge and will cause new problems
in practice.

Weak magnetic modulation \cite{Zhou} or even all-electric \cite{Popescu}
implementations \cite{Khod} for spin filters are more expected. Since spin
states can be manipulated efficiently through spin-orbit couplings (SOCs)
\cite{she}, one may carry out all-electric spin-based devices in SOC systems
without the need of external magnetic field and magnetic materials. For
example, when the transport occurs in multichannel regime, Rashba SOC \cite%
{Rash} in two-terminal quantum wires can polarize the electron beams \cite%
{Bulg}. This behavior may be an alternative route for all-electric
spin filters. However, for the view of the \textit{spintronic}
device performance, the single-channel devices are more desirable
because they suffer from much less spin relaxation \cite{Pram},
which does harm to most spintronic devices. At the same time,
single-channel sample is helpful for the miniaturization of the
functional elements in devices. Thus, we focus our attention on
one-dimensional quantum wires in the present work. Due to the
time-reversal symmetry, the SOC alone in single-channel wires proves
unable to generate any spin polarization \cite{Zhai2}, which means
that spin filtering can not be realized through only SOC mechanism
in single-channel wires. Because a weak magnetic field can break the
time reversal symmetry \cite{Sted}, it is full of possibility to
build spin filters in single-channel SOC systems with weak magnetic
modulation by designing certain models.

In the present work, we propose theoretical schemes for spin filters in
one-dimensional quantum wires through modulations of both magnetic
potentials and Rashba spin-orbit couplings. Two kinds of modulated
structures are considered. For case I, the magnetic potential is spatially
homogenous and the Rashba spin-orbit coupling is periodically modulated
along the wire. The periodic Rashba potentials result in two coinciding
energy gaps for spin-up and down electrons, while a weak magnetic potential
can break the time-reversal symmetry, separating the two gaps and therefore
inducing 100\% spin polarization within the two energy intervals. For case
II, the Rashba spin-orbit coupling is spatially uniform and the magnetic
potential is periodically modulated. Full spin polarization can be obtained
by the periodic magnetic potential. The more attractive result obtained in
this case is that not only the amplitude of the spin polarization but also
its sign can be changed conveniently by tuning the Rashba strength, i.e. a
negtive/positive switching effect for spin polarization is obtained.

\textit{\ }\textbf{Models and analysis:}\textit{\ }The geometries\textit{\ }%
we consider are two one-dimensional quantum wires with spatially
modulated magnetic potential and spin-orbit coupling illustrated in
Fig. 1. In Fig. 1(a), each periodic unit consists of one SOC segment
and one non-SOC segment with the same length of $a/2$ ($a$ is set at
20 nm in the following calculations). The magnetic field is applied
uniformly along the wire. In Fig. 1(b), the periodic unit consists
of one magnetic and one non-magnetic region also with the same
length of $a/2.$ The gate voltage is laid uniformly along the wire.
The symbols of $V_{g}$ in the figure indicate the applied gate
voltages to control the Rashba strengths, which are typically on the
order of 10$^{-11}$ eVm \cite{Nitt}. To provide a clear
illustration, we label the segments in series: $1,$ $2$ $...j,$
$j+1...$, as shown in Fig. 1.

\begin{figure}[tbph]
\begin{center}
\resizebox{8cm}{!}{\includegraphics*[183,542][424,706]{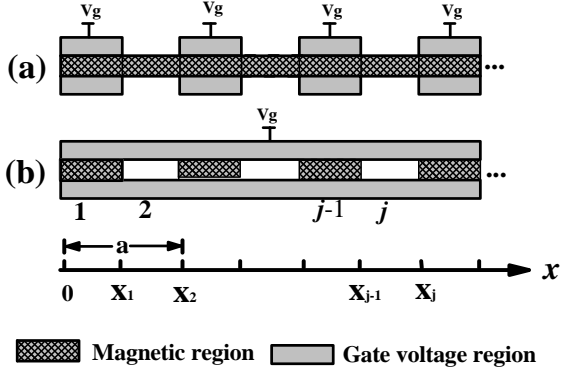}}
\end{center}
\caption{Schematic diagrams of two spatially modulated structures.
(a) The magnetic potential is uniform along the wire and the gate
voltages are periodically applied to control the Rashba SOCs. (b)
The gate voltage is uniform and the magnetic potentials are periodic
along the wire.}
\end{figure}

The Hamiltonians of the two modulation structures can be written as a
universal formula:%
\begin{equation}
H=\frac{p_{x}^{2}}{2m^{\ast }}+\sigma _{z}V_{0}g(x)-\dfrac{\alpha _{0}}{%
\hbar }\sigma _{z}p_{x}f(x),
\end{equation}%
where the effective mass of electrons $m^{\ast }$ is set as 0.067 $m_{e\text{
}}(m_{e}$ is the mass of the free electron), $p_{x}$ is the $x$-component of
the momentum operator, and $\sigma _{z}$ is the Pauli spin operator. The
denotation $V_{0}$ is a spin-dependent parameter for the strength of the
Zeeman-like potential. The parameter $\alpha _{0}$ indicates the strength of
SOC. Two functions $g(x)$ and $f(x)$ are introduced to describe the spatial
modulations of the magnetic potential and the SOC, respectively. For Fig.
1(a), $g(x)=1$ throughout the wire$,$ and $f(x)=1$ in the odd $j$ segments
and 0 in the even $j$ segments. For Fig. 1(b), $f(x)=1$ throughout the wire$%
, $ and $g(x)=1$ and 0 in the odd $j$ and even $j$ segments, respectively.

The wave functions $\Phi _{j}$ in odd $j$ segments and $\Psi _{j}$ in even $%
j $ segments can be written as:

\begin{subequations}
\begin{eqnarray}
\text{\ }\Phi _{j}=a_{j}e^{ik_{1}x}\left\vert \uparrow \right\rangle
+b_{j}e^{ik_{2}x}\left\vert \uparrow \right\rangle
+c_{j}e^{ik_{3}x}\left\vert \downarrow \right\rangle
+d_{j}e^{ik_{4}x}\left\vert \downarrow \right\rangle ,\text{ }j=1,3,5... \\
\text{ }\Psi _{j}=a_{j}e^{ik_{1}^{^{\prime }}x}\left\vert \uparrow
\right\rangle +b_{j}e^{ik_{2}^{\prime }x}\left\vert \uparrow \right\rangle
+c_{j}e^{ik_{3}^{\prime }x}\left\vert \downarrow \right\rangle
+d_{j}e^{ik_{4}^{\prime }x}\left\vert \downarrow \right\rangle ,\text{ }%
j=2,4,6....
\end{eqnarray}
The denotations $\left\vert \uparrow \right\rangle $ and $\left\vert
\downarrow \right\rangle $ express the eigenspinor states $\left(
\begin{array}{c}
1 \\
0%
\end{array}%
\right) $ and $\left(
\begin{array}{c}
0 \\
1%
\end{array}%
\right) .$ Suppose four wave vector functions $\tilde{k}_{1}(\alpha ,V),$ $%
\tilde{k}_{2}(\alpha ,V),$ $\tilde{k}_{3}(\alpha ,V),$ $\tilde{k}_{4}(\alpha
,V)$, and they are expressed as the following:

\end{subequations}
\begin{subequations}
\begin{eqnarray}
\tilde{k}_{1}(\alpha ,V) &=&\left( \alpha +\sqrt{\alpha ^{2}+\frac{2\hbar
^{2}(E-V)}{m^{\ast }}}\right) \frac{m^{\ast }}{\hbar ^{2}}, \\
\tilde{k}_{2}(\alpha ,V) &=&\left( \alpha -\sqrt{\alpha ^{2}+\frac{2\hbar
^{2}(E-V)}{m^{\ast }}}\right) \frac{m^{\ast }}{\hbar ^{2}}, \\
\tilde{k}_{3}(\alpha ,V) &=&\left( -\alpha -\sqrt{\alpha ^{2}+\frac{2\hbar
^{2}(E+V)}{m^{\ast }}}\right) \frac{m^{\ast }}{\hbar ^{2}}, \\
\tilde{k}_{4}(\alpha ,V) &=&\left( -\alpha +\sqrt{\alpha ^{2}+\frac{2\hbar
^{2}(E+V)}{m^{\ast }}}\right) \frac{m^{\ast }}{\hbar ^{2}},
\end{eqnarray}
where $E$ is the incident electron energy. Then for Fig.1(a), $\mathit{k}%
_{i}=\tilde{k}_{i}(\alpha _{0},V_{0})$ in Eq.(2a)$,$ and $\mathit{k}%
_{i}^{\prime }=\tilde{k}_{i}(0,V_{0})$ in Eq.(2b). For Fig.1(b), $\mathit{k}%
_{i}=\tilde{k}_{i}(\alpha _{0},V_{0})$ in Eq.(2a), and $\mathit{k}%
_{i}^{\prime }=\tilde{k}_{i}(\alpha _{0},0)$ in Eq.(2b). Using boundary
conditions \cite{Mats,Zuli,Yao} at the interfaces of SOC/non-SOC or
magnetic/non-magnetic, we can get the transfer matrix \cite{Shi, Gong1} for
the wave functions at $j$ and $j-1$ segments and obtain the outgoing spin
states.

Note that the injected current in our scheme is completely unpolarized,
which is theoretically simulated by a mixed quantum state of two orthogonal
spin states \cite{Niko}, for example, $\left(
\begin{array}{c}
1 \\
0%
\end{array}%
\right) $ and $\left(
\begin{array}{c}
0 \\
1%
\end{array}%
\right) $ are chosen in our calculation. For an incident electron in the
spin-up state $\left\vert \uparrow \right\rangle $, the conductance of the
outgoing spin-up state is $G^{\uparrow \uparrow }$ and that of the outgoing
spin-down state is $G^{\downarrow \uparrow }.$ Similarly, for an incident
electron in the spin-down state $\left\vert \downarrow \right\rangle $, the
conductances of the outgoing spin-up and spin-down states are $G^{\uparrow
\downarrow }$ and $G^{\downarrow \downarrow },$ respectively. The spin
polarization along $z$ axes is obtained from the formula \cite{Niko}:

\end{subequations}
\begin{equation}
P_{z}(E_{F})=\frac{G^{\uparrow \uparrow }+G^{\uparrow \downarrow
}-G^{\downarrow \downarrow }-G^{\downarrow \uparrow }}{G^{\uparrow \uparrow
}+G^{\uparrow \downarrow }+G^{\downarrow \downarrow }+G^{\downarrow \uparrow
}}.
\end{equation}
Here we would like to emphasize that any two orthogonal spin states with
equal probabilities can be used to simulate the unpolarized current. If we
choose orthogonal spin states $\left(
\begin{array}{c}
1 \\
0%
\end{array}%
\right) $ and $\left(
\begin{array}{c}
0 \\
1%
\end{array}%
\right) $ to simulate the unpolarized current, the conductance $%
G^{\downarrow \uparrow }=G^{\uparrow \downarrow }=0$, because $\left(
\begin{array}{c}
1 \\
0%
\end{array}%
\right) $ and $\left(
\begin{array}{c}
0 \\
1%
\end{array}%
\right) $ are eigenspinors of the Hamiltonian Eq.(1). Although spin-resolved
conductances $G^{\uparrow \uparrow }$, $G^{\downarrow \uparrow }$, $%
G^{\uparrow \downarrow }$, $G^{\downarrow \downarrow }$ rely on the the
orthogonal component of the injected unpolarized current, the spin
polarization does not.

\bigskip

\textbf{Results and Discussion:} Figure 2(a) shows the spin
polarization as a function of the Fermi energy for the modulation of
Fig. 1 (a), with the Rashba strength $\alpha _{0}$ is fixed at 0.04
a.u. ($1$ a.u.$=1.44\times 10^{-9}$ eVm). When no magnetic potential
is applied, the periodic SOC will induce coinciding energy gaps for
spin-up and -down electrons \cite{Gong1}. No spin polarization is
achieved because the gaps for spin-up and -down electrons have the
same position and width. The striking feature is that the
periodically modulated Rashba potential, even combining with a weak
magnetic modulation ($V_{0}=0.4$ meV), can induce appreciable spin
polarization. We get two energy intervals, within which 100\% spin
polarization with opposite signs are obtained. Increasing the
magnetic modulation to $V_{0}=2.0$ meV, the two gaps become farther
away from each other. In this scheme, the weak magnetic potential
plays the critical role to break the time-reversal symmetry. In
addition, we also investigate the case that $\alpha _{0}=0$ and
$V_{0}=2.0$ meV, no appreciable spin polarization is induced (not
shown). This demonstrates that the magnetic potential with the
magnitude $V_{0}=2.0$ meV is too small to be an effective
spin-selective barrier, yet it is sufficient to break the time
reversal symmetry.

\begin{figure}[tbph]
\begin{center}
\resizebox{8cm}{!}{\includegraphics*[230,407][479,668]{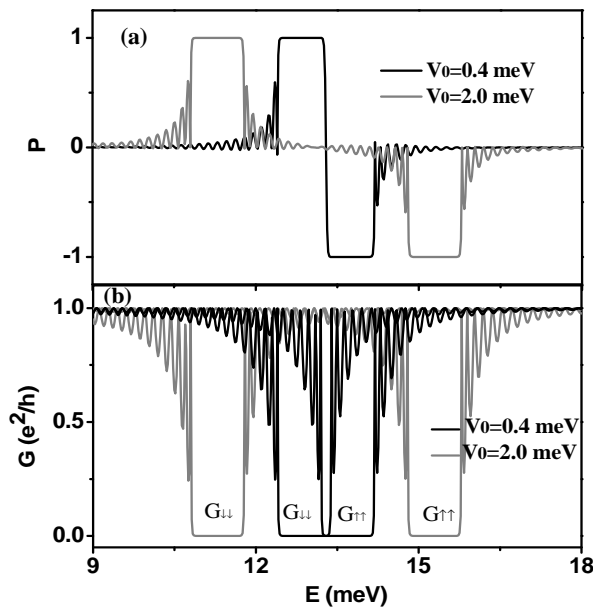}}
\end{center}
\caption{Results for the structure of Fig. 1(a). (a) The
spin-polarization as a function of the incident energy. (b) The
spin-dependent conductance as a function of the incident energy. The
magnetic potential $V_{0}$ is given as 0.4 meV and 2 meV. The Rashba
SOC $\protect\alpha_{0} $ is fixed at 0.04 a.u..}
\end{figure}

To better understand the behavior of the spin polarization in Fig. 2(a), we
investigate the spin-resolved conductance. Figure 2(b) shows the spin-up
conductance $G^{\uparrow \uparrow }$ \ and spin-down conductance $%
G^{\downarrow \downarrow }$ as a function of the incident energy. When a
weak magnetic potential is applied ($V_{0}=0.4 $ meV), gaps for spin-up and
-down are partially separated. Increasing the magnetic modulation to $%
V_{0}=2.0$ meV, the two gaps are completely separated. For the magnetic
modulation, spin-up and -down electrons see different magnetic potentials,
while for the Rashba modulation, they see the same Rashba potential. Both
the SOC and magnetic potential are indispensable for the spin filter effect:
the width of the gaps is determined by the periodic SOC strengths, while the
separation of the gaps depends on the application of the magnetic potential.

Figure 3(a) displays the spin polarization versus the incident
energy for the structure in Fig. 1(b) with the magnetic potential
parameter $V_{0}=1.0$ meV. When the Rashba strength $\alpha _{0}=0$,
the periodic magnetic modulations can result in 100\% spin
polarization in two energy intervals \cite{Zhou}. Here we pay more
attention to the influence of the uniform SOC. Applying a nonzero
SOC, we find that both the spin-up and -down gaps float toward lower
energy region. Comparing the curve corresponding to
\textquotedblleft $\alpha =0"$ and the one \textquotedblleft $\alpha
=0.032$ $a.u.",$ it is found that Rashba SOC can inverse the spin
polarization from 100\% to -100\% within certain energy ranges
(13.2$\sim $13.8 meV in the Fig. 3(a)). To clearly illustrate this
point, we plot Rashba SOC dependence of the spin polarization in
Fig. 3(b), in which energy is given as 13.5 meV. A transition from
positive to negative polarization is clearly seen, i.e. a
positive/negtive spin polarization switching effect is obtained by
tuning the SOC strength. The previous schemes for spin filters
generally just create polarized current out of an unpolarized
source. Here, we provide a more flexible property for spin filter.
In addition, we would like to emphasize that in Fig. 1(a) the
magnetic field just has the contribution to break the time reversal,
while in Fig. 1(b), it must bring an appreciable energy gap.
Quantitatively comparing Fig. 2 and Fig. 3(a), we believe that
modulation of Fig. 1(a) is more advantageous over that of Fig. 1(b),
if the lower magnetic potential is the main factor to be considered.

\begin{figure}[tbph]
\begin{center}
\resizebox{8cm}{!}{\includegraphics*[161,347][375,662]{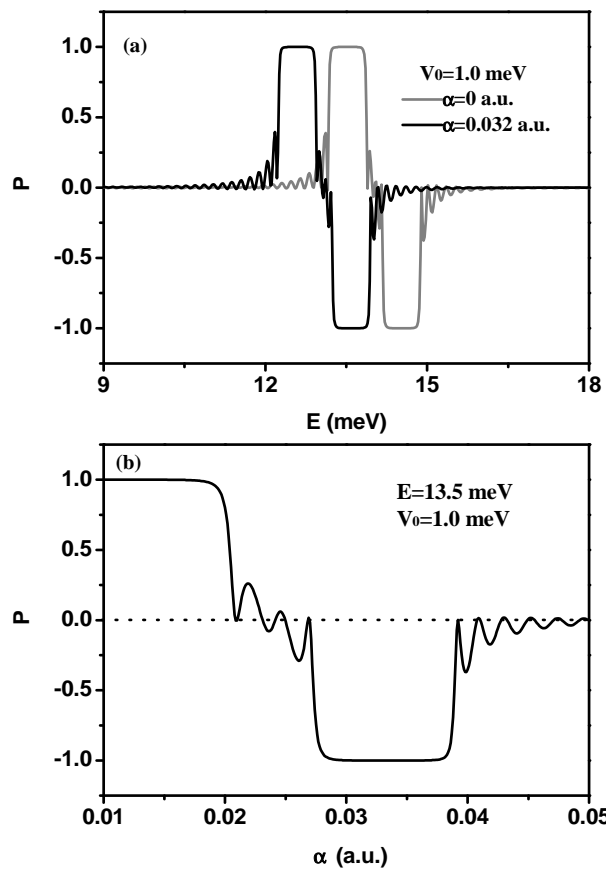}}
\end{center}
\caption{Results for structure of Fig. 1(b). (a) The
spin-polarization versus the incident energy. (b) The spin
polarization versus the Rashba spin-orbit coupling strength. }
\end{figure}

For the case that both the Rashba SOC and magnetic potential are
periodically modulated, some investigations have been conducted previously
\cite{Wang2}. For the same structure, we obtain different results compared
with the report of Ref.\cite{Wang2}. No spin polarization can be achieved
when the magnetic field is zero in our calculation, because the Rashba
spin-orbit coupling can not break the time reversal symmetry \cite{Zhai2}.
In Ref.\cite{Wang2}, however, even when the magnetic field is zero, they
obtain the full spin polarization. The critical difference results from that
the incident current with spin state $\left(
\begin{array}{c}
1 \\
1%
\end{array}%
\right) $ \ in Ref.\cite{Wang2} is only unpolarized along $z$ axes, but not
a completely unpolarized current. For this geometry, it is found that with
the application of the periodic SOC, the width of the positive polarization
is enlarged, while that of the negative polarization is suppressed, as shown
in Fig. 4. For the periodic magnetic modulation, although gaps for spin-up
and -down electrons have different positions, yet they have the same width
(see Fig. 3(a)), because the width of the gap is determined by the absolute
magnitude of the periodic potential. With the presence of the periodic
Rashba potential, the balance of the potential strengths for spin-up and
-down electrons is upset, which induces the disparity between the positive
and negative spin polarization observed in Fig. 4.

\begin{figure}[tbph]
\begin{center}
\resizebox{8cm}{!}{\includegraphics*[175,439][389,602]{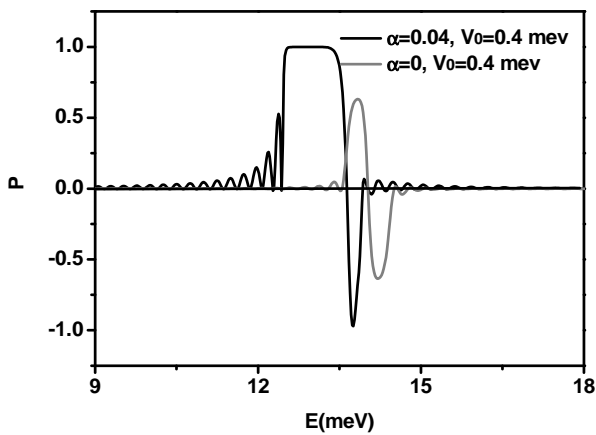}}
\end{center}
\caption{The spin polarization versus the incident energy with
periodic modulation of both the magnetic potential and the Rashba
SOC.}
\end{figure}

\textbf{Conclusion:} Theoretical schemes for spin filter are proposed
through investigations on the spin-dependent electron transport in
one-dimensional quantum wires with spatially modulated weak magnetic
potentials and Rashba spin-orbit couplings. Two kinds of modulation
structures are mainly investigated. For case I, by combining the periodic
SOC with the weak homogenous magnetic field and two separated gaps for
spin-up and -downs electrons are obtained, and therefore full spin
polarizations with opposite signs are realized. For case II, the periodic
magnetic potential results in two separated gaps, and the spin polarity
within the gaps can be switched by the uniform SOC modulation. The spin
filters are implemented through experimentally available Rashba interaction
and weak magnetic modulation. They may be useful in the future design of
spin-based devices.

\textbf{Acknowledgement:} This work was supported by\textbf{\ }the
National Natural Science Foundation of China with grant No 1067027,
the Grand Foundation of Shanghai Science and Technology (05DJ14003),
and 973-project under grant No. 2006CB921300.

\end{document}